# Growth and characterization of conducting LaAlO$_3$/EuTiO$_3$/SrTiO$_3$ heterostructures

G. M. De Luca, R. Di Capua, E. Di Gennaro, F. Miletto Granozio, and M. Salluzzo

*CNR-SPIN, and Dipartimento di Fisica Università "Federico II" di Napoli, Complesso Monte S. Angelo, Via Cinthia, I-80126 Napoli, Italy*

A. Gadaleta, I. Pallecchi, and D. Marrè

*CNR-SPIN and Dipartimento di Fisica, Università di Genova, Via Dodecaneso 33, I-14146 Genova, Italy*

C. Piamonteze, M. Radovic

*Swiss Light Source, Paul Scherrer Institut, CH-5232 Villigen PSI, Switzerland*

Z. Ristic, S. Rusponi

*Institute of Condensed Matter Physics, Ecole Polytechnique Fédérale de Lausanne, CH-1015 Lausanne, Switzerland*

**Abstract**
We studied the structural, magnetic and transport properties of LaAlO$_3$/EuTiO$_3$/SrTiO$_3$ heterostructures grown by Pulsed Laser Deposition. The samples have been characterized in-situ by electron diffraction and scanning probe microscopy and ex-situ by transport measurements and x-ray absorption spectroscopy. LaAlO$_3$/EuTiO$_3$/SrTiO$_3$ films show a ferromagnetic transition at T<7.5 K, related to the ordering of Eu$^{2+}$ spins, even in samples characterized by just two EuTiO$_3$ unit cells. A finite metallic conductivity is observed only in the case of samples composed by one or two EuTiO$_3$ unit cells and, simultaneously, by a LaAlO$_3$ thickness equal or above 4 unit cells. The role of ferromagnetic EuTiO$_3$ on the transport properties of δ-doped LaAlO$_3$/EuTiO$_3$/SrTiO$_3$ is critically discussed.

The discovery of a quasi two-dimensional electron system (Q2DES) at the interface between LaAlO$_3$ (LAO) and SrTiO$_3$ (STO) insulators [1] stimulated an intense research on novel functional oxide heterostructures. In particular, the intriguing electronic properties of the Q2DES in the LAO/STO system have generated an intense debate about the physics of oxide interfaces. Among the most interesting characteristics of the Q2DES we cite a transport dominated by a remarkably large Rashba spin-orbit coupling [2] and low temperature two-dimensional superconductivity (SC). Both phenomena can be tuned by using electric field effect. In particular a metal (SC at low T) to insulating transition can be obtained even at room temperature [3,4]. At the same time, different reports suggest the presence of some magnetic order, possibly coexisting with superconductivity [5-7]. In view of these unique properties, the integration of LAO/STO into modern electronics, and in particular in spintronics, is an exciting challenge. The recent reports of resonant amplification [8] and of long lifetimes [9] of spin-polarized electrons injected in STO heterostructures are encouraging in these directions. On the other hand, the realization of a spin-polarized Q2DES at the equilibrium remains at the moment only an ambitious target. Indeed, the interpretation of the experimental signatures of magnetism in LAO/STO heterostructures remains controversial, and more recent reports suggest only a modest spin polarization of the Q2DES, possibly related to oxygen vacancies [10].

A promising route to obtain a homogeneous and robust spin polarization at equilibrium in LAO/STO is by δ-doping with a magnetic layer. Yet, the first attempts at δ-doping have shown that in general it induces charge localization [11,12] and, in the case of insertion of rare-earth oxides, it gives rise to electron correlation effects possibly enhanced by structural distortions [13]. Here we show that EuTiO$_3$ (ETO) can be used as magnetic δ-doping layer to obtain heterostructures characterized by a Q2DES and a ferromagnetic order due ETO. film. This result represents a first step toward the realization of an oxide spin-polarized Q2DES.

ETO is a perovskite whose lattice parameters (a=b=c=3.905Å) perfectly match those of STO, allowing strain-free epitaxial growth. Its electronic structure is also very similar to that of STO, with a broad valence band originating from O–2p states and a conduction band formed by Ti–3d states. However, ETO possesses an additional narrow Eu–4f band, lying just below the Fermi level, located within a direct band gap <0.93eV, much smaller than those of STO (<3.2eV) and LAO (<5.6eV) [14]. Bulk ETO has a G-type antiferromagnetic ground state below the Néel temperature T$_N$~5.5 K, which is switched to a ferromagnetic state by doping [15] or lattice strain [16]. The lattice, electronic and magnetic properties of ETO, as well as the non-polar character of the (001) surface, similar to STO, make it a good candidate as magnetic dopant layer for LAO/STO interfaces.

We deposit LAO(n)/ETO(m)/STO (n=0-12 unit cells (uc), and m=0-5 uc) epitaxial heterostructures by Reflection High Energy Electron Diffraction (RHEED) assisted pulsed laser deposition (PLD) from sintered Eu$_2$Ti$_2$O$_7$ and crystalline LAO targets onto TiO$_2$ terminated (001) STO substrates. A KrF excimer laser (wavelength 248 nm, pulse rate 1 Hz) is focused on the target at a fluence of 1.3 J/cm$^2$. To prevent the formation of the competing phase Eu$_2$Ti$_2$O$_7$, ETO deposition must be carried out at oxygen partial pressure ($p_{O2}$) and deposition temperature ($T_d$) lower than those usually employed for LAO/STO heterostructures. We adopted two set of deposition parameters: Set A, $T_d$=650-700°C and $p_{O2}$=1x10$^{-4}$ mbar for both ETO and LAO films; Set B, $T_d$=600°C and $p_{O2}$=1x10$^{-7}$ mbar for the growth of ETO and

$T_d$=800°C and $p_{O2}$=1x10$^{-4}$ mbar for LAO. No annealing procedures were carried out after the deposition to avoid the recrystallization of ETO to the $Eu_2Ti_2O_7$ phase at higher oxygen pressures. The samples were in both cases slowly cooled down to room temperature, with a rate of 3°C/min, in $10^{-4}$ mbar of $O_2$.

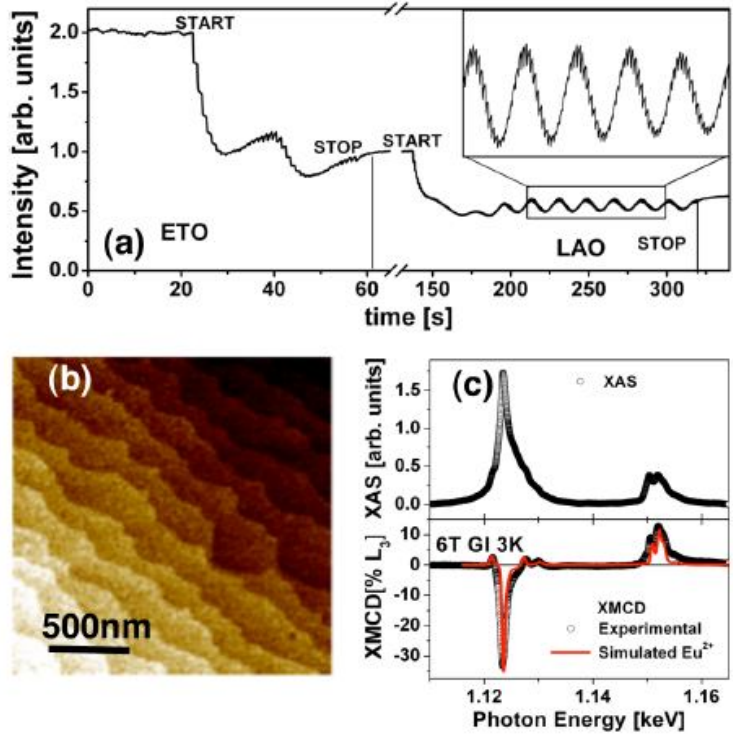

Figure 1 : (a) Specular RHEED intensity vs. deposition time during the growth of a LAO/ETO film on STO single crystal. (b) AFM topography on a conducting a LAO(10)/ETO(2)/STO sample.(c) upper panel: XAS sum-spectra of circular polarized(c+ and c-)spectra at 3 K and 6 T, with field almost parallel to the interface (grazing incidence, GI). Bottom panel: corresponding XMCD spectrum, which is the difference between the XAS data acquired with c+ and c- circular polarizations and corresponding simulated spectra (red line) using multiplet scattering theory for an $Eu^{2+}$ ion..

*In-situ* structural and surface characterizations were performed by RHEED (STAIB high pressure RHEED) and scanning probe microscopy (Omicron VT-AFM) in the Modular facility for Oxide Deposition and Analysis (MODA). Chemical, electronic and magnetic characterizations were carried out by Eu-$M_{4,5}$ edge X-ray Absorption Spectroscopy (XAS) and X-ray magnetic Circular Dichroism (XMCD) at the X-TREME beam-line of the Swiss Light Source at the Paul Scherrer institute [17]. Finally transport measurements, including Hall-effect, were performed using a Quantum Design PPMS system in the temperature range from 300K to 1.8K and magnetic fields up to 9 Tesla.

In Fig. 1 we present a series of experimental characterizations of the LAO/ETO/STO heterostructures. The RHEED analysis shows that all the samples have a layer-by-layer growth and that their structural quality is similar to LAO/STO films. Atomic force microscopy shows that the LAO/ETO/STO morphology mimics those of the STO substrates on which they are deposited, with stepped terraces separated by one-uc high step-edges. Finally, the XAS and XMCD spectra demonstrate that Eu is in $Eu^{2+}$ oxidation state with some admixtures of $Eu^{3+}$ (<30%) which can form as consequence of the oxidizing atmosphere during the deposition, or due to partial Eu/La substitution.

The saturated magnetic moment, calculated using the sum rules, are 3.3±0.2 $\mu_B$/Eu 3.5±0.2 $\mu_B$/Eu and 3.7±0.2 $\mu_B$/Eu for 1uc, 2uc and 4uc ETO thin films respectively, therefore being a slowly increasing function of the thickness. These magnetic moments are lower than the expected value of 7 $\mu_B$/Eu of $Eu^{2+}$ ions (S=7/2). The deviation is related to the presence of non-magnetic $Eu^{3+}$ and to the extremely reduced ETO thickness in our heterostructures. In Fig. 2 we show typical magnetic hysteresis loop at 3K and a temperature dependence of the maximum of the XMCD signal of a LAO/ETO/STO heterostructure characterized by only 2uc ETO δ-doping layer embedded between 10uc LAO film and the STO single crystal. The data show that thin ETO films are ferromagnetic with a low field remanence of 0.3 $\mu_B$/Eu (inset of Fig 2a). A full alignment of the spin moments is achieved above 1 Tesla. The ferromagnetic transition temperature, measured in 0.1 Tesla, is about 7.5 K. The switching from the bulk-like antiferromagnetic character to the ferromagnetic behavior of the ETO layers could be related to electron doping and/or a slight oxygen excess as reported for films grown by PLD in oxidizing atmosphere [18].

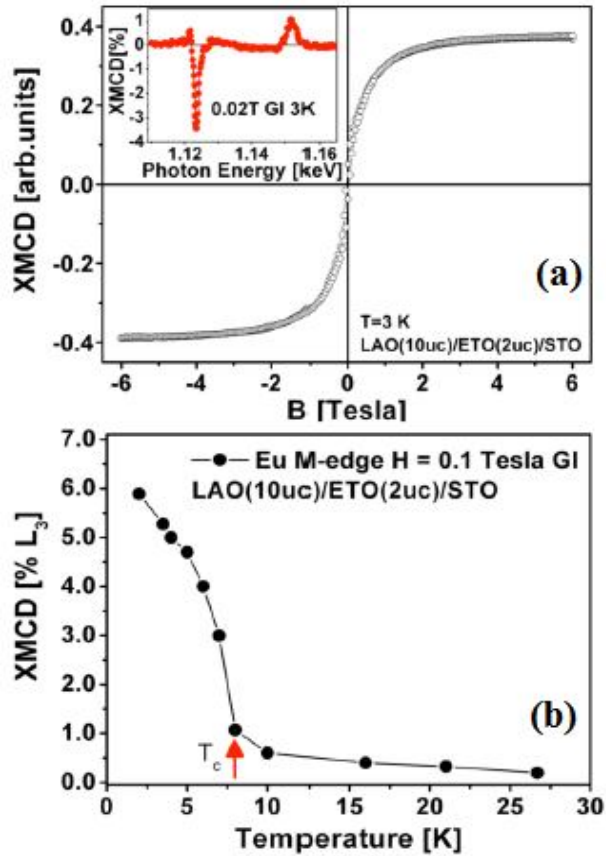

Figure 2 : (a) Magnetic hysteresis loop at 3K in grazing incidence conditions for a metallic LAO(10)/ETO(2)/STO heterostructure. The inset shows an XMCD spectra at 0.02 Tesla and 3K. (b) Temperature dependence of the XMCD intensity normalized to $L_3$ measured at 0.1 T after saturating the magnetization at 6 T at low temperatures. The red arrow indicates the ferromagnetic transition temperature.

We now examine the electric transport behavior of these heterostructures. It is well known that LAO/STO interfaces become conducting for a LAO thickness equal or larger than the critical value of 4uc, which is usually explained in terms of polar catastrophe mechanism.
Consequently, we first explored the effect of the ETO thickness on the conductance of heterostructures having a number of LAO layers larger than the critical value of 4uc (namely 10uc). We find that the number of ETO layers is a crucial parameter. As shown in Fig.3a, a finite conductance is present only for ETO films of one or two unit cells, whereas for higher thicknesses (up to 6uc) the conductance of the interface is negligible in the limit of our experimental set up. On the other hand, by fixing the ETO thickness to one or two unit cells, we show in Fig.3b that the conductance is finite only for LAO thicknesses equal or larger than the critical value of 4uc, as in the case of usual LAO/STO interfaces. This is expected in a polar catastrophe scenario since the ETO layers are non polar and do not alter the polarity stacking of the atomic layers. It is worth noting that STO/ETO/STO heterostructures, where 10uc of STO replaces the LAO film on top of ETO, are insulating, suggesting a crucial role of LAO in the creation of a metallic interface.

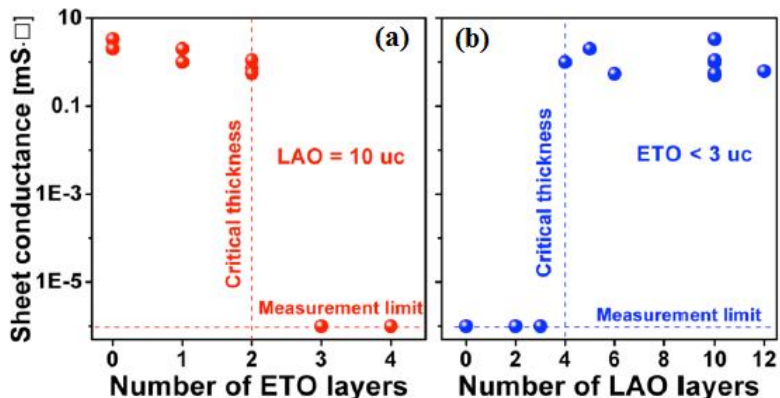

Figure 3: Sheet conductance measured at room temperature of LAO/ETO/STO samples (a) as function of the number of ETO layers (LAO thickness > 4uc) and (b) as function of the number of LAO layers (ETO thickness < 3uc).

The sheet charge carrier density of LAO/ETO/STO heterostructures is similar to that one of LAO/STO samples and ranges from $3-5 \cdot 10^{13}$ (Set A) to $8-10 \cdot 10^{13}$ e-/cm$^{-2}$ (Set B) low temperatures. The temperature dependence of the sheet resistance and of the mobility for some representative samples is shown in Fig.4a,b. The sheet resistance exhibits a metallic behavior with values below 30 kΩ at 300 K typical of LAO/STO oxide interfaces. On the other hand, LAO/ETO/STO samples exhibit, below <50K, a significant resistance upturn. This resistance upturn is also sometimes observed in LAO/STO heterostructures at reduced doping, and it is generally attributed to weak localization. However, in LAO/ETO/STO any attempt to fit the low temperature data using standard weak localization models [19] fails.

The Hall mobility of LAO/ETO/STO samples in Fig.4b shows the characteristic power law temperature dependence similar to standard LAO/STO as consequence of scattering by optical phonons [20]. However, at low temperature, it shows a downturn below 50 K corresponding to the upturn of the sheet resistance. Thus, the transport characterization indicates that the ETO layer suppresses the conductance and the mobility of the interface introducing another source of scattering in the system. Enhanced scattering associated with the presence of an additional interface has to be excluded, since LAO(10)/STO(2)/STO heterostructures grown in the same conditions, and characterized by a 2uc STO film replacing ETO, show transport properties similar to standard LAO/STO [Fig. 4a, blue closed circles].

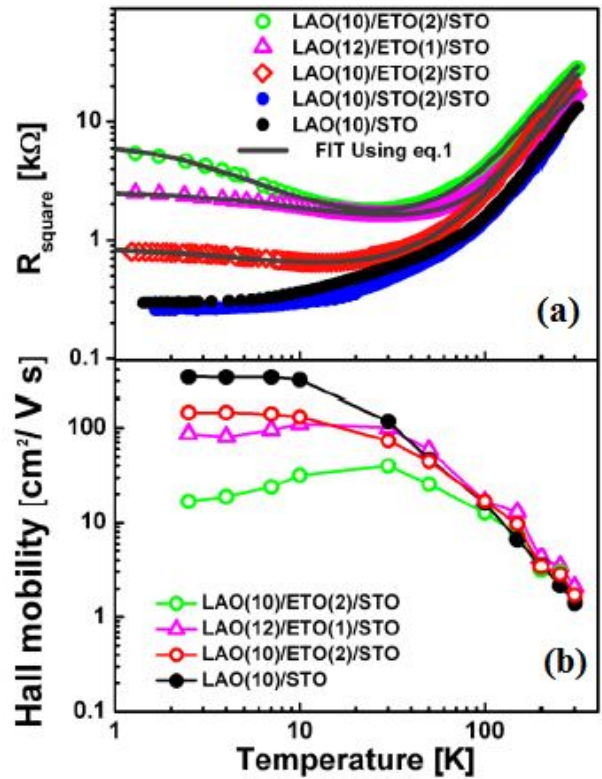

Figure 4: (a) Temperature dependence of the sheet resistance and (b) of the mobility of LAO/ETO/STO and LAO/STO heterostructures as labeled in the figure: Continuous black lines in (a) are the fitting curves obtained using eq. 1 and eq.2.

Thus, the characteristic upturn of the low temperature resistivity and the simultaneous mobility suppression have to be attributed to the specific role of ETO. We propose that the mechanism for such mobility suppression is Kondo-like magnetic scattering of the Q2DES. To verify the consistency of this hypothesis, we fit the transport data taking into account the presence of Kondo-scattering. The complete formula used to fit the sheet resistance curves is assumed to be the sum of four terms:

$$R = R_0 + R_K(T) + AT^2 + R_{BG}(T) \qquad (1)$$

$R_0$ is the temperature independent term due to non magnetic impurity scattering, the $T^2$ contribution is a consequence of a quasi-non-retarded interaction between dressed quasi-particles in STO [21], $R_{BG}(T)$ is the phonon scattering term described by the Bloch-Grüneisen law and $R_K(T)$ is the Kondo term related to scattering by magnetic impurities. For $R_K(T)$ we use the zero field generalized Hamann expression [22]:

$$R_K = C\left(1 - \frac{\ln(T/T_K)}{\sqrt{(\ln(T/T_K))^2 + \pi^2(S(S+1))}}\right) \qquad (2)$$

Here, $T_K$ is the effective Kondo temperature and S is the effective spin of the magnetic scattering centers. We get a consistent fit (Fig.4a, continuous black lines) using the Kondo model for all the LAO/ETO/STO heterostructures. In particular we find a Kondo temperature which varies slightly around $T_K \approx 10K$, while the parameter S is always found to be close to S=0.22, in good a agreement with the value expected from numerical renormalization group theory for a Kondo impurity with spin 1/2 and similar to experimental values obtained on other titanates [23].

In a polar catastrophe scenario, electrons are expected to be transferred to ETO layers first. The large energetic cost associated to on-site coulomb repulsion of two electrons in 4f Eu sites (U=6 eV from ref. [14]) hinders a filling of the Eu 4f states. Thus one should expect a filling of the Ti3d states belonging to the ETO layer. However, as long as the ETO film is thin enough, at least a fraction of the doping electrons could leak also into STO at the interface leading to formation of a Q2DES as in standard LAO/STO. On the other hand, when the ETO thickness exceeds 2uc, all the electrons remain trapped in ETO and apparently do not contribute to the formation of a Q2DES extending into the underneath STO. This accounts for the observation that the mobile charge carrier density, the sheet resistance and the overall characteristics of LAO/ETO/STO interfaces, excluded the enhanced low temperature scattering, are similar to that of standard LAO/STO interfaces. It remains to be explained why the electrons transferred to few unit-cells thick (3-6) ETO are not mobile. One possibility is that electron correlations, possibly favored by a partial overlapping of Ti-3d and Eu-4f unoccupied states [14], play an important role in the observed behavior.

This qualitative picture can also account for the enhanced scattering, of magnetic origin, observed in the case of LAO/ETO/STO heterostructures. Indeed, while we cannot establish at the moment the origin of the Kondo impurity scattering centers, one possible candidate is an excess of electrons transferred to localized titanium states at the interface and acting as dilute magnetic scattering centers. In this picture the ETO layer is a source of localized magnetic scattering centers located at the interface. Another explanation is that a fraction of $Eu^{2+}$ ions, possibly replacing some $Sr^{2+}$ in STO, act as Kondo impurities. However, the ferromagnetic character of ETO is hardly reconciled with a disordered intermixed $Eu_{1-x}Sr_xTiO_3$ layer at the interface, which has been shown to be antiferromagnetic (with reduced $T_N$) or not magnetically ordered [24].

The results presented in this paper show that ETO is a suitable δ-doping layer for the realization of artificial Ferromagnetic/Q2DES system based on LAO/STO oxide heterostructures. The Kondo-like low temperature transport behavior suggests that the Q2DES interact with spin 1/2 magnetic moment, presumably associated to electrons trapped in interfacial Ti-3d states. Thus, for the range of doping investigated, the Q2DES does not show a ferromagnetic-metallic character, which would manifest in a metallic low temperature transport and eventually a downturn of the resistivity at the ETO ferromagnetic transition. Nevertheless, other experimental investigations are needed to establish whether the Kondo-like low transport behavior can be tuned to ferromagnetic metallic character as expected for a spin-polarized Q2DES.

In conclusion we have grown high quality LAO/ETO/STO heterostructures using RHEED assisted PLD. This system shows a Q2DES when the ferromagnetic ETO layer is below 3 uc. The results show that δ–doping by magnetic layers is a viable route to engineering the properties of the Q2DES in LAO/STO heterostructures.

We acknowledge funding from the Ministero dell'Istruzione, dell'Università e della Ricerca (grant PRIN 2010-11 – OXIDE).